\documentclass{JHEP3}
\usepackage{amsmath}
\usepackage{amsfonts}
\usepackage{graphicx}
\usepackage{epsfig}

\newcommand{\acc}{AdS/CFT correspondence }
\newcommand{\DDir}{\relax{D\kern-.7em{/}}}

\newcommand{\haf}{\frac{1}{2}}
\newcommand{\inv}[1]{\frac{1}{#1}}

\newcommand{\RR}{\mathbb{R}}

\newcommand{\ra}{\rightarrow}

\newcommand{\soo}{\Rightarrow}
\newcommand{\oso}{\Leftrightarrow}

\newcommand{\be}{\begin{equation}}
\newcommand{\ee}{\end{equation}}
\newcommand{\bea}{\begin{equation*}}
\newcommand{\eea}{\end{equation*}}

\newcommand{\abs}[1]{\left\vert#1\right\vert}

\newcommand{\ave}[1]{\left\langle #1\right\rangle}

\renewcommand{\pr}{\partial}

\newcommand{\nin}{\relax{\in\kern-.8em{/}}}

\newcommand{\Te}{\Theta}
\newcommand{\te}{\theta}
\newcommand{\al}{\alpha}
\newcommand{\bt}{\beta}

\newcommand{\lm}{\lambda}
\newcommand{\de}{\delta}
\newcommand{\De}{\Delta}

\newcommand{\Om}{\Omega}

\newcommand{\sig}{\sigma}

\newcommand{\trm}{\textrm}

\newcommand{\tbt}{\tilde \bt}
\newcommand{\bx}{\bar x}

\newcommand{\hh}{\tilde h}

\title {Non-Local Effects of Multi-Trace Deformations in the AdS/CFT
Correspondence}

\preprint{\hepth{0504177}\\WIS/08/05-APR-DPP}
\author{Ofer~Aharony, Micha~Berkooz and
Boaz~Katz\\\\
Department of Particle Physics, Weizmann Institute of Science,
Rehovot 76100, Israel\\ \\
{\tt E-mail:
Ofer.Aharony@weizmann.ac.il, berkooz@wisemail.weizmann.ac.il,
boaz.katz@weizmann.ac.il} }

\abstract{The AdS/CFT correspondence relates deformations of the
CFT by ``multi-trace operators'' to ``non-local string theories''. The
deformed theories seem to have non-local interactions in the
compact directions of space-time; in the gravity approximation the
deformed theories involve modified boundary conditions on the
fields which are explicitly non-local in the compact directions.
In this note we exhibit a particular non-local property of the
resulting space-time theory. We show that in the usual backgrounds
appearing in the AdS/CFT correspondence, the commutator of two
bulk scalar fields at points with a large enough distance between
them in the compact directions and a small enough time-like
distance between them in AdS vanishes, but this is not always true
in the deformed theories. We discuss how this is consistent with
causality.}
\begin{document}
%

%
\section{Introduction}

The anti-de Sitter (AdS) / conformal field theory (CFT) correspondence
\cite{Maldacena:1997re,Gubser:1998bc,Witten:1998qj} and its
generalizations to non-conformal field theories relate local field
theories to specific solutions of string theory (or M theory). In many
cases the string (M) theory has a limit where it is well-approximated
by a specific background of ten (eleven) dimensional supergravity. The
dimension of the space in this background is larger than the dimension
of the space that the local field theory lives on; generally there are
some additional compact directions, and one additional ``radial''
direction which roughly corresponds to the scale of the field theory.

In the limit where the string theory is well-approximated by gravity,
it is approximately local in the higher dimensional space (at least at
low energies, and at distances much larger than the string scale).
This locality is far from manifest in the original field theory, and
it is not even clear which precise property of the field theory it
corresponds to. The locality in the ``radial'' direction seems to
indicate some sort of decoupling between phenomena occurring at
different length scales (even when they are at the same position in
space), and the locality in the compact directions implies specific
relations between the correlation functions of operators corresponding
to Kaluza-Klein harmonics in these compact directions. Note that the
relation between the causal properties of the dual theory in the
gravity approximation and those of the local field theory is relatively
well-understood \cite{Horowitz:1999gf,Kabat:1999yq,Gregory:2000an}
(see \cite{Hubeny:2005qu} for a recent discussion).
However, the relation between locality and causality is non-trivial
when space is not flat and infinite.

One way to try to understand better the emergence of bulk locality
is to try to find deformations that will break it, and that will
lead to field theories which are dual to string theories with
non-local bulk physics. A way to do this was proposed in
\cite{Aharony:2001pa}. The AdS/CFT correspondence maps string
theory vertex operators, which are related to fields moving in the
corresponding background, to specific operators in the dual field
theory; when the dual theory is a large $N$ gauge theory, these
operators (for low angular momenta on the compact manifold)
are single-trace operators, and we will use this
nomenclature in general for the operators which are dual to single
bulk fields. A deformation of the field theory by such a
single-trace operator corresponds to introducing a source at the
boundary of anti-de Sitter space for the corresponding bulk field
\cite{Gubser:1998bc,Witten:1998qj}, which is a local effect. A
deformation of the field theory action by multi-trace
deformations, involving a product of two or more of the
single-trace operators (at the same point in space-time) results
in a seemingly non-local theory on the AdS side.

Such deformations were first discussed in \cite{Aharony:2001pa},
where the deformed theories were called ``Non-Local String
Theories''. One reason for this name was that such deformations
are manifestly non-local on the string worldsheet of the dual
string theory, because they involve a product of vertex operators
each of which is integrated over the full string worldsheet. A
second reason for this name was that these deformations seem to be
non-local also in space-time, because generally they involve a
product of fields on AdS space, each of which arises from some
Laplacian eigenfunction on the compact coordinates. In the
original paper \cite{Aharony:2001pa} it was not clear whether the
deformation is local in the AdS coordinates or not, but this was
later clarified in
\cite{Witten:2001ua,Berkooz:2002ug,Sever:2002fk} where it was
shown that the effect of the deformation on the bulk fields can be
described by a modified boundary condition on the boundary of AdS
space, so it is manifestly local in the AdS coordinates. However,
it still seems to be non-local in the additional compact
coordinates.

One manifestation of the non-locality, which was discussed in
\cite{Aharony:2001pa,Aharony:2001dp}, is that the force between
D-branes localized in the compact directions seems to grow faster
than allowed by locality. In this paper we describe another
manifestation of the non-locality. We will work in the
approximation where the bulk fields are well-described by a weakly
coupled field theory on a curved space $AdS_{d+1}\times M$, and
fluctuations of the background can be ignored; of course, this is
the only case where locality (and causality) properties have a
clear meaning. We focus on the simple example of a deformation
involving the modes of ten dimensional (or eleven dimensional in M
theory) free scalar fields, though we expect the generalization to
other cases to be straightforward (we have explicitly generalized our
results to free vector fields). In a local bulk theory we expect
the commutator of scalar fields at two points to vanish whenever
the geodesic connecting the two points is space-like and there is
no time-like geodesic connecting the points\footnote{
In cases where there are both space-like and time-like geodesics,
one expects the contribution of the time-like geodesics to the
commutator to be non-zero. Such a situation can occur, for instance,
whenever there are compact cycles,
since then one can always go around them any number of times to
obtain space-like geodesics between any pair of points.}, since
one would expect the space-like geodesic to dominate the path
integral in such a case. In flat space this requirement follows
from causality, but this is not true in spaces of the form
$AdS_{d+1}\times M$. We will show in section \ref{review} that in such
spaces causality generally allows commutators of scalar fields not
to vanish in particular pairs of points (described in section
\ref{review}), which are connected by causal curves even though the only
geodesics connecting them are space-like.

We will then show in section \ref{non deformed} that for the standard boundary
conditions on AdS space the commutators of scalar fields vanish at
such pairs of points, as expected in a local bulk theory. On the
other hand, we will show in section \ref{Double trace deformation}
that for the modified
boundary conditions corresponding to multi-trace deformations the
commutators no longer vanish at such pairs of points, even when their
space-like geodesic distance is arbitrarily large. This gives a precise
manifestation of the bulk non-locality in such theories. Even though we
will exhibit this phenomenon only for a special case involving scalar
fields in anti-de Sitter space, we expect it to
be completely general for arbitrary multi-trace deformations (including
those of non-conformal theories).

We will start by reviewing some properties of $AdS_{d+1}$ and the \acc
in section \ref{review}, focusing on the causal properties of spaces
of the form $AdS_{d+1}\times M$. We will show that there are pairs of
points which are causally connected even though the only geodesics
connecting them are space-like.  In section \ref{non deformed} we will
show that the commutator of a free massive scalar field in the
non-deformed theory (with standard boundary conditions) on
$AdS_{d+1}\times M$ vanishes for such pairs of points (this is
actually shown only for even $d$). In section \ref{Double trace
deformation} we will calculate the correction to the propagator due to
a specific double-trace deformation to first order, and show that the
commutator is non-vanishing for some of the relevant
pairs of points. We summarize our
results in section \ref{summary}. Two appendices contain some technical
details.

\section{Causality and locality in $AdS_{d+1}\times M$ and the
\acc}\label{review}

\subsection{Review of $AdS_{d+1}$ in global coordinates}

In this paper we will discuss bulk theories living on $d+1$ dimensional
anti-de Sitter (AdS) space in global coordinates
\begin{equation} \label{metrictautheta}
ds^2=\frac{R^2}{\cos^2\theta}(-d\tau^2+d\theta^2+\sin^2\theta~d\Om_{d-1}^2),
\end{equation}
where $d\Om_{d-1}^2$ is the metric on $S^{d-1}$ and
$0\leq\te\leq\frac\pi2$, which are dual
\cite{Maldacena:1997re,Gubser:1998bc,Witten:1998qj} to conformal
field theories (CFTs) on $S^{d-1}\times \RR$. The vector
$\pr_\tau$ is a globally defined time-like Killing vector, so
$\tau$ serves as a global time coordinate in the bulk (which is
identical near the boundary to the time coordinate of the dual
CFT). It will be useful to define an `origin' of the coordinate
system \eqref{metrictautheta} at $\tau=\theta=0$. Since
$AdS_{d+1}$ is homogeneous there is of course nothing special
about this `origin'.

AdS space has some unusual features which will play an important role in
this work. All geodesics starting at the origin have constant
position on $S^{d-1}$ and are distinguished by their $\te(\tau)$ behavior.
All time-like geodesics reach the points $\te=0,\tau=k\pi$ after
proper time equal to $k\pi$. Not every two points on AdS which can
be connected by a causal curve can be connected by a causal
geodesic. The region $G$ of AdS space which is
reached by causal geodesics leaving the
origin is shown in figure \ref{penrose2}(a). A
time-like geodesic connecting the origin to a point in $G$
is shown in figure \ref{penrose2}(b). Any point in $G$ is
related to a point on the curve $\te=0$ by an $SO(d,2)$ isometry that
doesn't affect the origin.

\setlength{\unitlength}{1mm}
\begin{figure}[!h]
\begin{picture}(270,140)
\newsavebox{\figaa}
\savebox{\figaa}(100,130)[bl]{
\put(40,40){\line(0,1){100}} \put(60,40){\line(0,1){100}}
\put(40,40){\line(1,1){20}}\put(60,60){\line(-1,1){20}}
\put(40,80){\line(1,1){20}}\put(60,100){\line(-1,1){20}}
\put(40,120){\line(1,1){20}}
\put(33,30){$\te=0$} \put(57,30){$\te=\haf\pi$}
\put(25,40){$\tau=0$}
\put(25,80){$\tau=\pi$}\put(40,40){\circle*{1}}\put(37,37){$x_2$}
\put(61,60){$\haf\pi$}
\put(25,120){$\tau=2\pi$}\put(61,100){$\frac 32\pi$}
\put(45,60){G} \put(45,100){G}\put(45,135){G}}

\newsavebox{\figbb}
\savebox{\figbb}(100,130)[bl]{
\put(40,40){\line(0,1){100}} \put(60,40){\line(0,1){100}}
\put(40,40){\line(1,1){20}}\put(60,60){\line(-1,1){20}}
\put(40,80){\line(1,1){20}}\put(60,100){\line(-1,1){20}}
\put(40,120){\line(1,1){20}}
\put(33,30){$\te=0$} \put(57,30){$\te=\haf\pi$}
\put(25,40){$\tau=0$}
\put(25,80){$\tau=\pi$}\put(40,40){\circle*{1}}\put(37,37){$x_2$}
\put(61,60){$\haf\pi$}
\put(25,120){$\tau=2\pi$}\put(61,100){$\frac 32\pi$}
\put(47.5,100){\circle*{1}}\put(49,102){$x_1$}
\qbezier(40,40)(55,60)(40,80) \qbezier(40,80)(47,90)(47.5,100)
\thicklines \put(47.5,60){\vector(0,1){0.5}}
 }

\put(-15,-20){\usebox{\figaa}}\put(32,0){(a)}
\put(60,-20){\usebox{\figbb}}\put(108,0){(b)}
\end{picture}
\caption{Diagrams of AdS space. The dependence on the $\Om_{d-1}$
coordinates is suppressed. The regions (marked by G) in which
points can be connected to the origin by a time-like geodesic are
shown in (a). A time-like geodesic in AdS connecting the origin
with a point which is also in the `boundary affected'
 region (see below) is
shown in (b).\label{penrose2}}
\end{figure}
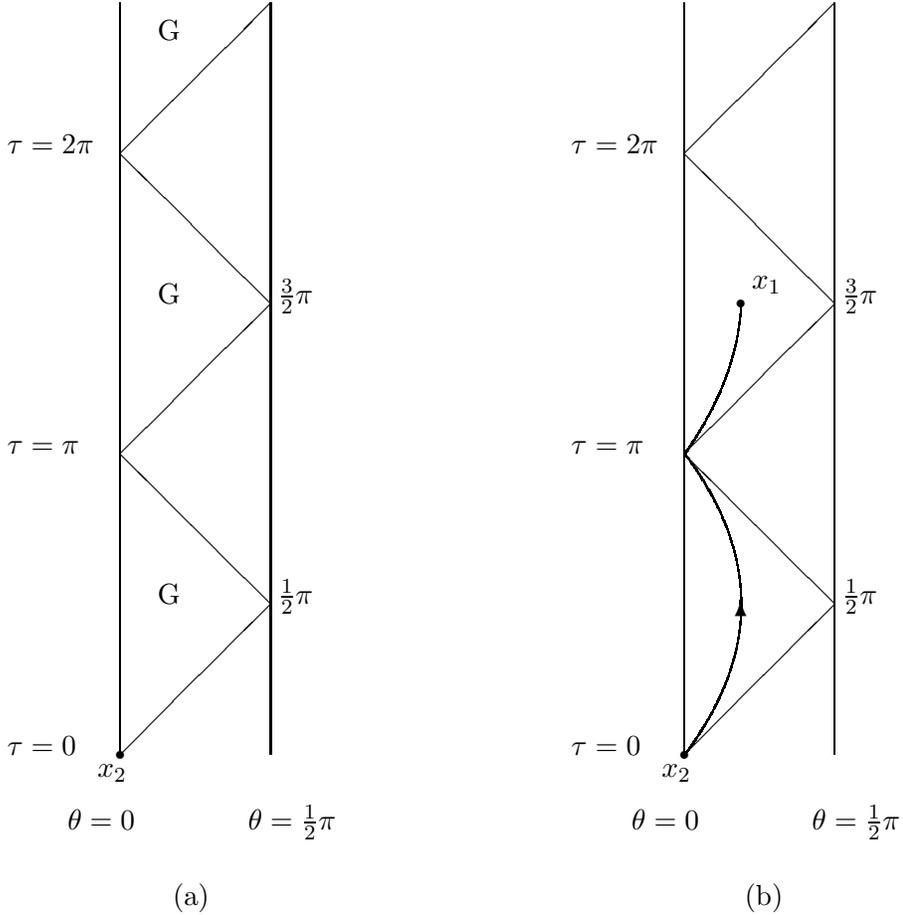

AdS space has a boundary at $\te=\pi/2$, and a theory on this space
depends on the boundary conditions there. A point $x_1$ for which there is
enough time for a null geodesic starting at the origin to reach
the boundary and go from there to $x_1$, namely:
\begin{align}\label{BCA} \tau_1\geq\pi-\te_1,
\end{align}
will be called `boundary affected'. A point for which this is not
possible will be said to be `boundary unaffected' (see figure
\ref{penrose}(a)). In order for the boundary conditions to affect
the commutator of two fields at $x_1$ and at $x_2=0$, $x_1$ must be in the
boundary affected region (hence the name). Boundary affected
points can be connected to the origin by a causal curve with
arbitrarily long proper time. Such a curve is shown in figure
\ref{penrose}(b).

\setlength{\unitlength}{1mm}
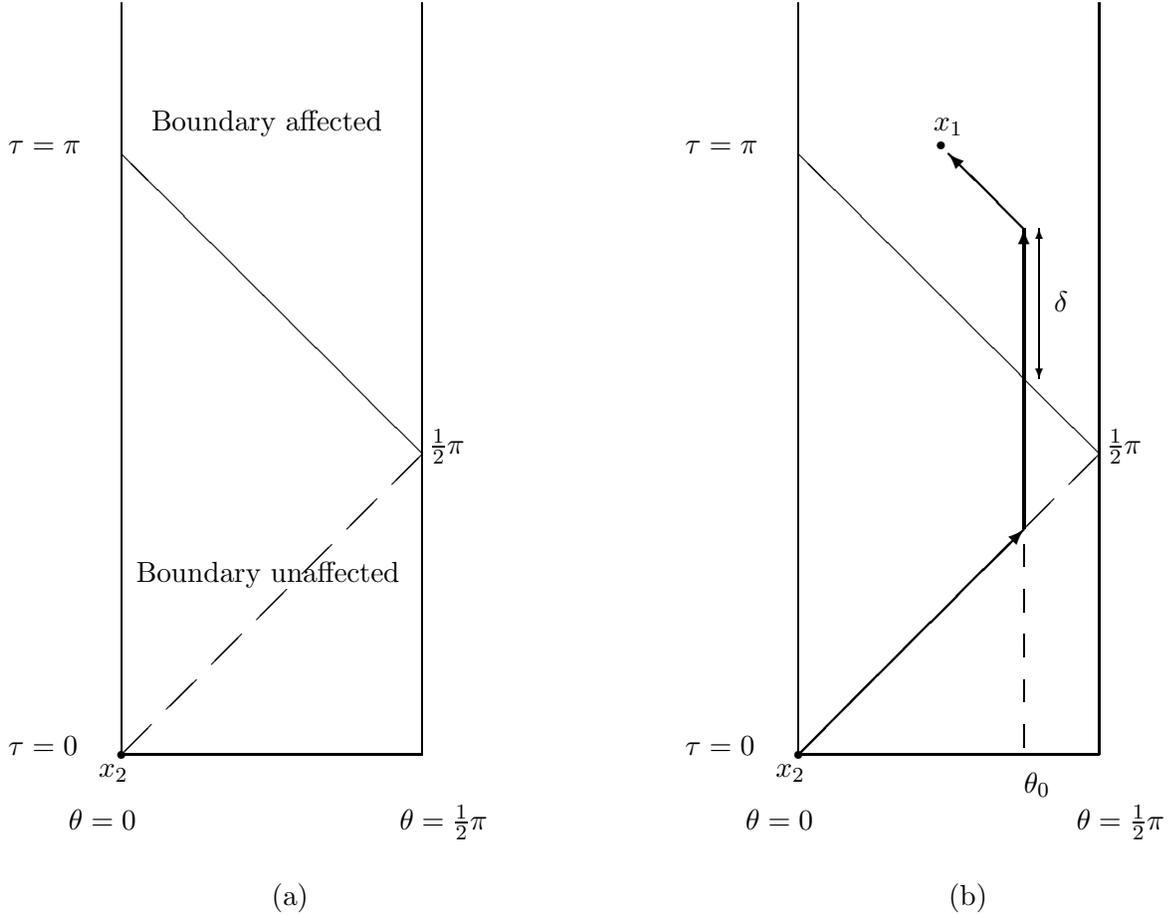
\begin{figure}[!h]
\begin{picture}(270,140)
\newsavebox{\figa}
\savebox{\figa}(100,130)[bl]{
\put(40,40){\line(0,1){100}} \put(80,40){\line(0,1){100}}
\put(40,40){\line(1,0){40}}
\multiput(40,40)(6,6){7}{\line(1,1){4}}
\put(80,80){\line(-1,1){40}}
\put(33,30){$\te=0$} \put(77,30){$\te=\haf\pi$}
\put(25,40){$\tau=0$}
\put(25,120){$\tau=\pi$}\put(40,40){\circle*{1}}\put(37,37){$x_2$}
\put(81,80){$\haf\pi$}
\put(44,123){Boundary affected} \put(42,63){Boundary unaffected}}

\newsavebox{\figb}
\savebox{\figb}(100,110)[bl]{
\put(40,40){\line(0,1){100}} \put(80,40){\line(0,1){100}}
\put(40,40){\line(1,0){40}}
\multiput(40,40)(6,6){7}{\line(1,1){4}}
\put(80,80){\line(-1,1){40}}
\thicklines
\put(40,40){\vector(1,1){30}}\put(70,70){\vector(0,1){40}}
\put(70,110){\vector(-1,1){10}}\put(59,121){\circle*{1}}\put(58,123){$x_1$}
 \thinlines
\put(72,110){\vector(0,-1){20}}\put(72,90){\vector(0,1){20}}\put(74,99){$\de$}
\multiput(70,68)(0,-6){5}{\line(0,-1){3}}
\put(33,30){$\te=0$} \put(77,30){$\te=\haf\pi$} \put(70,35){$\te_0$}
\put(25,40){$\tau=0$}
\put(25,120){$\tau=\pi$}\put(40,40){\circle*{1}}\put(37,37){$x_2$}
\put(81,80){$\haf\pi$}}

\put(-30,-20){\usebox{\figa}}\put(30,0){(a)}
\put(60,-20){\usebox{\figb}}\put(120,0){(b)}
\end{picture}
\caption{Diagrams of AdS space. The distinction between regions for
which the commutator is potentially affected or unaffected by the boundary
conditions is shown in (a). A causal curve in AdS connecting the
origin with a point in the `boundary affected' region is shown in
(b). The proper time along this curve is larger than
$\de/\cos(\te_0)$. By taking $\te_0$ to $\pi/2$ it can be made
arbitrarily long.\label{penrose}}
\end{figure}

Rotating  $\tau$ to $\tau_E=-i\tau$, the metric
\eqref{metrictautheta} becomes the Riemannian metric
\begin{equation} \label{metrictauEtheta}
ds^2=\frac{R^2}{\cos^2\te}(d\tau_E^2+d\theta^2+\sin^2\te~d\Omega_{d-1}^2).
\end{equation}
This is $EAdS_{d+1}$, the maximally symmetric space with Euclidean
signature and negative curvature.
Any two points on $EAdS$ can be connected by a geodesic.
Any two sets
of two points with the same geodesic distance between them are
related by an isometry. As in Lorentzian AdS space, we can
define the `origin' as $\tau_E=\theta=0$.

\subsection{A brief review of the AdS/CFT correspondence}

The \acc states that string theories on
$AdS_{d+1}\times M$ are dual to conformal field theories in $d$
dimensions ($CFT_d$). Fields on AdS space (which could come from some
Kaluza-Klein mode of a higher dimensional field) are mapped by the
correspondence to a specific class of local primary operators in the CFT
called ``single-trace operators''; when the dual theory is a large $N$
gauge theory with adjoint fields these are the operators which may be
written as a single trace of a product of fields (more precisely, this is
only true for fields which do not carry large angular momenta in $M$).
The states created by these operators and
their descendants are mapped to single-particle states in the bulk, while
states created by ``multi-trace operators'' which are products of such
operators are multi-particle states in the bulk.

In this paper we will focus on scalar fields in the bulk.
A massive scalar field $\phi$ with mass squared $m^2$ on
$AdS_{d+1}$ is related to a scalar
operator $\cal O$ with conformal dimension $\De$ in the $CFT_d$, with
\begin{align}\label{de to m}
\De=\frac d2\pm\sqrt{\frac {d^2}4+R^2m^2}
\end{align}
(the sign is always positive when $\frac {d^2}4+R^2m^2>1$, and can be
either positive or negative otherwise).
A general solution to the homogeneous Klein-Gordon equation of
motion of such a field
behaves near the boundary as :
\begin{align}\label{KG decomposition}
\phi=\al(\bx)(\cos(\theta))^{d-\De}(1+\trm{O}(\cos^2(\theta)))+\
\bt(\bx)(\cos(\theta))^{\De}(1+\trm{O}(\cos^2(\theta)))
\end{align}
with $\bx$ a point on the boundary of AdS.

The standard boundary condition, before we
perform any deformations, is given by $\al(\bx)=0$.
In the undeformed theory the expectation value of the operator ${\cal O}$ is
related to the value of $\beta$ -- more precisely it is given by
$\tbt$ defined
as \cite{Klebanov:1999tb}:
\begin{align}
\tbt(\bx)=(2\De-d)\bt(\bx).
\end{align}
A single-trace deformation of the CFT by subtracting from the
action the term $\int d^d\bx h(\bx) {\cal O}(\bx)$ is described on
AdS by the modified boundary condition $\al(\bx) = h(\bx)$. As in
any field theory, correlation functions of the operator ${\cal
O}(\bx)$ may be computed by taking functional derivatives with
respect to $h(\bx)$.

It is also possible to deform the CFT action by multi-trace
operators \cite{Aharony:2001pa}. This type of deformation can also
be described by modifying the boundary conditions\footnote{In order
to compute correlation functions in some cases, depending on
the branch, the AdS/CFT correspondence formula might require some
modification
\cite{Muck:2002gm,Gubser:2002vv,Minces:2002wp}.}
\cite{Witten:2001ua,Berkooz:2002ug,
Sever:2002fk}. In the case of $n$ scalar fields
$\phi_1,\cdots,\phi_n$ corresponding to operators ${\cal
O}_1,\cdots,{\cal O}_n$, a deformation $W[{\cal
O}_1(\bx),\cdots,{\cal O}_n(\bx)]$ of the CFT action, where $W$ is a
functional of $n$ scalar functions on $\pr_{AdS_{d+1}}$, will
correspond to the modified boundary condition
\cite{Witten:2001ua,Sever:2002fk}
\begin{align}\label{general boundary condition}
&\al_i(\bx_0)=\frac {\de W[\tbt_1(\bx),\cdots,\tbt_n(\bx)]}{\de
\tbt_i(\bx_0)},~~~i=1,\cdots,n.
\end{align}

\subsection{General considerations of locality and causality on $AdS_{d+1}
\times M$}

In this paper we wish to study the effect of a multi-trace deformation of
the CFT on the corresponding fields in $AdS_{d+1}\times M$. We will argue
that after such a deformation the bulk theory is still causal, but it is
no longer local in a sense that we will explain below.

The metric on $AdS_{d+1}\times M$, using coordinates $y^i$ and a metric
$g_M$ on $M$, is:
\begin{align}\label{metric on AdSM}
ds^2=\frac{R^2}{\cos^2\theta}(-d\tau^2+d\theta^2+\sin^2\theta~d\Om_{d-1}^2)+
g_{Mij}dy^idy^j.
\end{align}
In the following, $x$ will
represent a point in $AdS_{d+1}$ and $y$ a point in $M$.
Consider two scalar fields
$\Phi_1(x,y)$, $\Phi_2(x,y)$ on $AdS_{d+1}\times M$
(the generalization to non-scalar
fields is straightforward). In order to study the locality and causality
properties of the theory on this space
we will be interested in their `bulk to bulk' commutator:
\begin{align}
C_{ij}(x_1,y_1;x_2,y_2)=\ave{[\Phi_i(x_1,y_1),\Phi_j(x_2,y_2)]}.
\end{align}
Obviously, the commutator should vanish whenever the points
$(x_1,y_1)$ and $(x_2,y_2)$ are not connected by a causal curve.

A multi-trace deformation of the CFT involving operators that
correspond to fields $\phi_i$ arising from
Kaluza-Klein (KK) modes of $\Phi_1(x,y)$, $\Phi_2(x,y)$
on $M$, will
correspond to changing the boundary conditions of these KK modes
according to \eqref{general boundary condition}. For multi-trace
deformations, unlike the case of single-trace deformations, the change in the
boundary conditions of specific KK modes according to
\eqref{general boundary condition} is a function of (other)
KK modes, and thus in general it cannot be written locally on $M$.
So, it looks like the theory with the new boundary conditions is non-local
in $AdS_{d+1}\times M$ (even though the boundary conditions are local in
$AdS_{d+1}$).

Obviously, the deformed CFT is still causal, so we expect that the
new boundary conditions cannot produce any non-causality in the bulk.
This is indeed the case, essentially because the global time
component of the metric $g_{\tau\tau}$ in \eqref{metric on AdSM}
goes to infinity at the boundary while the distance on $M$ stays
constant, so non-locality on $M$ at the boundary cannot produce
any non-causality. To see this explicitly, consider two points
$(x_1,y_1)$, $(x_2,y_2)$ on $AdS_{d+1}\times M$. Choose the
coordinate system on $AdS_{d+1}$ such that $x_2$ is at the
`origin' $x_2=0$. As explained above, the point
$x_1$ in $AdS_{d+1}$ has to satisfy exactly one of the two following
conditions (see figure \ref{penrose}):
\begin{itemize}
\item  The point $x_1$ is in the boundary unaffected region, namely the
time difference of $x_2$ and $x_1$ is not enough for a light
signal to leave $x_2$, go to the boundary, and come back to $x_1$ on
$AdS_{d+1}$.
In this case, no matter where $y_1$ and $y_2$ are on $M$, the new
boundary conditions cannot affect the commutator of fields at $(x_1,y_1)$,
$(x_2,y_2)$ (we are using the fact that the propagation is causal except
for possible effects of the new boundary conditions).
In particular, if $(x_1,y_1)$ and $(x_2,y_2)$ are not causally
connected, the commutator in the deformed theory
vanishes as for the undeformed theory.

\item There is a causal curve in $AdS_{d+1}$ with arbitrarily long
proper time connecting $x_2$ and $x_1$.
In this case, whatever the distance between $y_1$ and $y_2$ is on
$M$ (denote this distance $l_M(y_1,y_2)$), there is a causal curve
on $AdS_{d+1}$ connecting $x_2$ and $x_1$ with proper time greater
than $l_M(y_1,y_2)$. This curve can be lifted to a causal curve
connecting $(x_1,y_1)$ and $(x_2,y_2)$ on $AdS_{d+1}\times M$.
\end{itemize}
So, it is clear that even after the deformation the commutator will
remain zero for any two points on $AdS_{d+1}\times M$
which are not causally connected.

Next, we wish to study whether the deformed theory on $AdS_{d+1}\times M$
behaves according to our expectations from a local theory on $AdS_{d+1}
\times M$. Consider two points,
$(x_1,y_1)$ and $(x_2,y_2)$, on $AdS_{d+1}\times M$
(again assume $x_2$ is at the origin), such
that $x_1$ and $x_2$ are connected in AdS space by a time-like geodesic, and
such that the distance between $y_1$ and $y_2$ on $M$ is larger than the
geodesic time interval between $x_1$ and $x_2$ on AdS. This means that
on $AdS_{d+1}\times M$, $(x_1,y_1)$ is connected to $(x_2,y_2)$ by a
space-like geodesic. If in addition $x_1$ is in the `boundary
affected' region, then there is also a causal curve between the two
points as explained above, but there is no causal geodesic connecting
the two points.
Such points are shown in figure \ref{penrose2}(b) and are discussed
in more detail below. For such points, we would naively expect
that the commutator should vanish, since we would expect that in a
local theory the
space-like geodesic should dominate the path integral. In the next
section we will show that this expectation is indeed valid before
we perform the multi-trace deformation~--
the commutator of scalar fields in the
undeformed theory does indeed vanish for such points, at least for
scalar fields in odd-dimensional $AdS$ spaces (we have also proven this
for vector fields). In section \ref{Double trace deformation}
we will show that after the multi-trace deformation
(at least for the specific deformation described there) the commutator
no longer vanishes at these points, giving an indication of the non-local
nature of the deformed theory.

Let us describe in more detail the pairs of points that we are
interested in. In order to have
a time-like geodesic between $x_2$ and $x_1$,
$x_1$ must be in the region $G$ shown in figure \ref{penrose2}(a),
namely
\begin{align} k\pi +\te_1<\tau_1<(k+1)\pi-\te_1 \end{align}
for some integer $k$. In order for $x_1$ to also be in the `boundary
affected region' shown in figure \ref{penrose}, we must have
$k\geq 1$ so $\tau_1>\pi$. The proper time along the geodesic
connecting these points is then greater than $R\pi$. As we explained,
we are interested in pairs of points $y_1,y_2$ on $M$ such that
the distance $l_M(y_1,y_2)$ between them is
larger than this proper time; in particular it needs to be greater than
$R\pi$. It is not always possible to find such pairs of points.
For example, in the
$AdS_5\times S^5$ case the maximal distance between two points on $S^5$
is $R\pi$, since in this case the radii of $AdS_5$ and $S^5$ are equal,
so it is not possible. However, in other cases like M theory on $AdS_4\times
S^7$ or type II string theory on $AdS_3\times S^3\times T^4$
(with a large $T^4$), the compact space is large enough and we can find
such pairs of points. These are the theories we will be interested in
for the purposes of this paper.

The specific example that we will focus on in the rest of this paper is
a double-trace deformation of the CFT,
\begin{align}\label{DTD}
S_{CFT}\ra S_{CFT}-h\int {\cal O}_1(\bx){\cal O}_2(\bx)d^d\bx,
\end{align}
where the scalar operator ${\cal O}_i$ (of dimension $\Delta_i$)
corresponds to a scalar field $\phi_i$ on $AdS_{d+1}$, which arises
from the KK expansion of the field $\Phi^i$ on
$AdS_{d+1}\times M$. As discussed above, this deformation
corresponds to the following boundary conditions on $AdS_{d+1}$
\eqref{general boundary condition}:
\begin{align}\label{DTBC}
\al_1(\bx)&=h\tbt_2(\bx)=h(2\De_2-d)\bt_2(\bx), \cr
\al_2(\bx)&=h\tbt_1(\bx)=h(2\De_1-d)\bt_1(\bx).
\end{align}

We will be interested in the properties of the bulk to bulk
propagator when we impose the boundary conditions \eqref{DTBC}.
For simplicity we assume that the fields $\Phi^1$ and $\Phi^2$
both have a positive mass squared, and we will work in the
approximation in which they are free (it should be possible to
perturbatively add interactions between the fields as well). This
is not a physical case since it leads to the deformation
\eqref{DTD} being irrelevant, but it is the simplest case and we
expect the results to carry over in a straightforward manner also
to cases where the deformation is marginal or relevant (such cases
necessarily involve non-scalar fields on $AdS_{d+1}\times M$). The
fields $\Phi^i$ may be decomposed as:
\begin{align}\label{KK decomposition}
\Phi^i(x,y)=\sum_I\phi^i_I(x)Y_I(y),~~~~i=1,2,
\end{align}
where $Y_I$ are normalized eigenfunctions of of the Laplacian on
$M$, with eigenvalues $\lm^2_I$,
\begin{align}
\nabla^2_MY_I=-\lm^2_IY_I.
\end{align}
The zero mode is the constant function on $M$
\begin{align}
Y_0(y)=\inv{\sqrt{V_M}},
\end{align}
with $V_M$ the volume of $M$. The fields $\phi^i_I$ are the KK modes
of $\Phi^i$ on $AdS_{d+1}$, with masses squared $m^2_{iI}$ given by
\begin{align}
m^2_{iI}=m_i^2+\lm^2_I.
\end{align}
Without loss of generality we will focus on the case where
the $AdS_{d+1}$ fields $\phi_1$ and $\phi_2$ are the zero
modes of $\Phi^1$ and $\Phi^2$,
\begin{align}
\phi_1\equiv \phi^1_0,~~~~\phi_2\equiv\phi^2_0.
\end{align}

We are interested in computing the bulk to bulk Feynman propagator
\begin{align}
G^{ij}_{AdS\times
M}(x_1,y_1;x_2,y_2)=\ave{T(\Phi^i(x_1,y_1)\Phi^j(x_2,y_2))},~~~~~~i,j=1,2,
\end{align}
and more specifically the bulk to bulk commutator
\begin{align}
C^{ij}_{AdS\times
M}&(x_1,y_1;x_2,y_2)=\ave{[\Phi^i(x_1,y_1),\Phi^j(x_2,y_2)]}=\cr
&=i(\Te(\tau_1-\tau_2)-\Te(\tau_2-\tau_1))\trm{Im}(G^{ij}_{AdS\times
M}(x_1,y_1;x_2,y_2)).
\end{align}
Inserting the decomposition \eqref{KK decomposition}, we have:
\begin{align}
G^{ij}_{AdS\times
M}(x_1,y_1;x_2,y_2)&=\sum_{I,J}\ave{T(\phi^i_I(x_1)\phi^j_J(x_2))}
Y_I(y_1)Y_J(y_2)\cr
&\equiv \sum_{I,J}G^{ij}_{AdS~IJ}(x_1,x_2)Y_I(y_1)Y_J(y_2).
\end{align}
Without the deformation, the fields $\phi^i_I$ are all independent
so we have:
\begin{align}
G^{ij}_{AdS~IJ}(x_1,x_2)=\de^{ij}\de_{IJ}G_{AdS}(x_1,x_2;m^2_{iI}),
\end{align}
and we get for the non-deformed theory:
\begin{align}\label{non deformed prop}
&G^{ij}_{AdS\times
M}(x_1,y_1;x_2,y_2)=\de^{ij}\sum_IG_{AdS}(x_1,x_2;m^2_{iI})
Y_I(y_1)Y_I(y_2)\cr
&\soo C^{ij}_{AdS\times
M}(x_1,y_1;x_2,y_2)=\de^{ij}\sum_IC_{AdS}(x_1,x_2;m^2_{iI})Y_I(y_1)Y_I(y_2).
\end{align}
With the new boundary conditions \eqref{DTBC}, the correlation
function of the zero modes changes and the change in the
propagator is:
\begin{align}
&\De G^{ij}_{AdS\times M}(x_1,y_1;x_2,y_2)=\De
G^{ij}_{AdS}(x_1,x_2)Y_0(y_1)Y_0(y_2)\cr &=\inv{V_M}\De
G^{ij}_{AdS}(x_1,x_2)\cr &\soo\De C^{ij}_{AdS\times
M}(x_1,y_1;x_2,y_2)=\inv{V_M}\De C^{ij}_{AdS}(x_1,x_2),
\end{align}
where $G^{ij}_{AdS}, C^{ij}_{AdS}$ are the Feynman propagator
and commutator of the zero modes:
\begin{align}
&G^{ij}_{AdS}\equiv G^{ij}_{AdS~00}=\ave{T(\phi_i\phi_j)},\cr
&C^{ij}_{AdS}\equiv\ave{[\phi_i,\phi_j]}.
\end{align}
In particular, since the propagator without the deformation is
diagonal in $i$ and $j$, the off-diagonal term in the deformed theory
is simply given by
\begin{align}\label{its enough}
G^{12}_{AdS\times M}(x_1,y_1;x_2,y_2)&=\inv{V_M}
G^{12}_{AdS}(x_1,x_2)\soo,\cr C^{12}_{AdS\times
M}(x_1,y_1;x_2,y_2)&=\inv{V_M} C^{12}_{AdS}(x_1,x_2).
\end{align}

From the general considerations about the effect of the boundary
conditions on the commutator, we know that for $x_2$ at the origin
and $x_1$ in the `boundary unaffected' region $\tau_1<\pi-\te_1$,
there is no change in the commutator as a consequence of the
deformation.
We are interested in
points in the special region we mentioned above,
namely points in the boundary affected region which can be connected
to the origin by a geodesic
such that the geodesic distance of $y_1,y_2$ on M is larger than
the geodesic time interval on $AdS$. For such points we claim that
in the undeformed (local) theory the commutators vanish.
Obviously the off-diagonal commutators vanish in this case
so this reduces to the claim that
\begin{align}
C^{ii}_{AdS\times M}=0
\end{align}
for these points.
In the next section we will prove this for even $d$
($AdS_3,~AdS_5,~AdS_7,\cdots$) and for a general compact manifold $M$. We
further claim that the deformed theory no longer has this
property. By \eqref{its enough} it is enough to show that
$C^{12}_{AdS}(x_1,x_2)\neq 0$ for these points (with the boundary
condition \eqref{DTBC}). We will show this
in section \ref{Double trace deformation}. Thus, we will show that the
deformed theory behaves differently from the original local
theory.

\section{The commutator in the non-deformed theory}\label{non
deformed}

In this section we will prove, for even values of $d$, the following claim:
for a general compact manifold $M$, and for a pair of points on
$AdS_{d+1}\times M$ for which the projected points on $AdS_{d+1}$
are connected by a time-like geodesic, and for which the distance
between the projected points on $M$ is larger than the geodesic time
interval between the projected points on $AdS_{d+1}$, the
commutator of a free massive scalar field vanishes.

We believe that the claim is also true for odd values of $d$, and that
it is true for non-scalar fields as well; we have generalized the proof
to the case of vector fields with even $d$, but we will present here only
the proof for scalar fields.
The proof will involve relating
the propagators on AdS space to propagators in flat space.

\subsection{$AdS_3\times M$}

We begin by discussing the case of $AdS_3$.
The scalar Feynman propagators on $AdS_3$ and on $\RR^{2,1}$ for some given
mass $m$ are, respectively, \eqref{prop3}:
\begin{align}
&G_{AdS_3}(\mu;m^2)=i\inv{4\pi
R}\frac{e^{i\mu\sqrt{m^2+\inv{R^2}}}}{\sin\frac\mu R},\cr
&G_{\RR^{2,1}}(\mu;m^2)=i\inv{4\pi}\frac{e^{i\mu m}}{\mu},
\end{align}
where $\mu$ is the geodesic proper time between the two points
in each of these spaces. There is
a simple relation between the propagators (with different masses) :
\begin{align}\label{RA prop relation}
G_{AdS_3}(\mu;m^2)=\frac{\frac\mu R}{\sin(\frac\mu
R)}G_{\RR^{2,1}}(\mu;m^2+\inv{R^2}).
\end{align}
The Feynman propagator for a scalar particle with mass $m$ on $AdS_3\times M$
is\footnote{In general, the sum in \eqref{prop decomposition} is
not convergent in Lorentzian signature. Using the Euclidean
versions of \eqref{RA prop relation} and \eqref{proprel2} and then
analytically continuing back to Lorentzian signature, the results
\eqref{EAM prop} and \eqref{EAd prop} can be proven more
rigorously.} (using the decomposition \eqref{KK decomposition})
\begin{align}\label{prop decomposition}
G_{AdS_3\times M}(\mu,y_1,y_2;m^2)=\sum_I
G_{AdS_3}(\mu;m^2+\lm^2_I)Y_I(y_1)Y_I(y_2).
\end{align}
Thus, we can relate the propagator of a field on $AdS_3\times M$ to
the propagator on $\RR^{2,1}\times M$~:
\begin{align}\label{EAM prop}
G_{AdS_3\times M}(\mu,y_1,y_2;m^2)&=\frac{\frac\mu
R}{\sin(\frac\mu R)}\sum_I
G_{\RR^{2,1}}(\mu;m^2+\inv{R^2}+\lm^2_I)Y_I(y_1)Y_I(y_2)=\cr
&=\frac{\frac\mu R}{\sin(\frac\mu R)}G_{\RR^{2,1}\times
M}(\mu,y_1,y_2;m^2+\inv{R^2}),
\end{align}
and the commutator obeys
\begin{align}\label{commutators}
C_{AdS_3\times M}(\mu,y_1,y_2;m^2)=\frac{\frac\mu
R}{\sin(\frac\mu R)}C_{\RR^{2,1}\times
M}(\mu,y_1,y_2;m^2+\inv{R^2}).
\end{align}
Obviously, if we have points on $\RR^{2,1}\times M$
for which the distance on $M$ is larger than the proper time difference
$\mu$ on $\RR^{2,1}$, the commutator on $\RR^{2,1}\times M$ will vanish, since
these points are not causally connected. Using (\ref{commutators})
we find that for such points
the commutator on $AdS_3\times M$ will vanish also, proving the claim
for the case of $AdS_3$. Note that this
proof applies whenever the mass squared on $\RR^{2,1}\times M$ is non-negative,
namely for any $m^2 \geq -1/R^2$; this is precisely the allowed range
by the Breitenlohner-Freedman bound (a scalar with mass
squared $m^2$ in $AdS_{d+1}$ satisfies $m^2 \geq -d^2/4R^2$)
\cite{Breitenlohner:1982jf}.

\subsection{Higher odd dimensions}

For higher dimensions there is no simple correspondence between
the propagator in $AdS_{d+1}$ and in $\RR^{d,1}$. However, there is
a relation connecting the propagators on AdS for different
spacetime dimensions, derived in appendix \ref{propagator_in_AdS}
\eqref{Aprop rel}:
\begin{align}\label{proprel2}
G_{AdS_{d+3}}(\mu;m^2)\propto \inv{\sin(\frac\mu
R)}\frac{d}{d\mu}G_{AdS_{d+1}}(\mu;m^2+\frac{d+1}{R^2}).
\end{align}
Note that the Breitenlohner-Freedman
bound restricts the masses on both sides of
\eqref{proprel2} in the same way:
\begin{align}
m^2_{AdS_{d+3}}\geq -\frac{(d+2)^2}{4R^2}\oso m^2_{AdS_{d+1}}+
\frac{d+1}{R^2}\geq -\frac
{d^2}{4R^2}.
\end{align}
The factor of proportionality in \eqref{proprel2} is real and does
not depend on the mass (the normalization of the propagator is
defined so that the Klein-Gordon operator will give a delta
function for coinciding points, and the behavior at very small
distances does not depend on the mass). So, for a general compact
manifold $M$ we have by the same reasoning as above
\begin{align}\label{EAd prop}
G_{AdS_{d+3}\times M}(\mu;y_1,y_2;m^2)\propto \inv{\sin\frac\mu
R}\frac{d}{d\mu}G_{AdS_{d+1}\times
M}(\mu;y_1,y_2;m^2+\frac{d+1}{R^2}).
\end{align}
Since the factors of proportionality are real, we get:
\begin{align}\label{commutatorsdims}
C_{AdS_{d+3}\times M}(\mu;y_1,y_2;m^2)\propto \inv{\sin\frac\mu
R}\frac{d}{d\mu}C_{AdS_{d+1}\times
M}(\mu;y_1,y_2;m^2+\frac{d+1}{R^2}).
\end{align}
Since we know from the previous subsection
that for $AdS_3\times M$ the commutator vanishes for
$\mu<l_M(y_1,y_2)$ for any (allowed) mass value, we can deduce by
(\ref{commutatorsdims}) that the commutator for any odd dimension
will vanish there. Thus, we have proven our claim
for scalar fields on $AdS_{d+1}$ for all positive even values of $d$.

\section{The commutator in the deformed theory}\label{Double trace
deformation}

\subsection{Double-trace deformation}

In this section we consider the commutator of two scalar fields
$\Phi_1$ and $\Phi_2$ on $AdS_{d+1}\times M$ with the deformation
described at the end of section \ref{review}, which modifies the
boundary conditions of their zero modes according to equation
\eqref{DTBC}.
In the undeformed theory the commutator of $\Phi_1$ and $\Phi_2$ vanished,
and after the deformation it is given by
\begin{align}\label{Big com cor}
\ave{[\Phi_1(x_1,y_1),\Phi_2(x_2,y_2)]}=\frac{1}{V_M}\ave{[\phi_1(x_1),\phi_2(x_2)]}.
\end{align}
In particular, because we have chosen $\phi_1$ and $\phi_2$ to be the
zero modes, this commutator is independent of $y_1$ and $y_2$; for a
different choice of KK modes we will have some product of harmonics
appearing on the right-hand side of \eqref{Big com cor}, but for a generic
pair of points on $M$ this product will not vanish, regardless of their
geodesic distance. Thus, in order to see if the commutator vanishes
for the pairs of points we are interested in, we need to check if
the $AdS_{d+1}$ commutator:
\begin{align}
\ave{[\phi_1(x_1),\phi_2(x_2)]}
\end{align}
vanishes when $x_1$ and $x_2$ are connected by a time-like
geodesic with proper time larger than $R\pi$. This computation
involves only $AdS_{d+1}$ without any reference to $M$, using the
deformed boundary conditions \eqref{DTBC}.

In order to show that the commutator is not zero it is enough to
show that it is non-zero to leading order in $h$. We will first
compute the propagator in Euclidean space and then continue it to
Lorentzian space.
To first order in $h$, the correction to the propagator in
Euclidean signature due to the
boundary conditions \eqref{DTBC} is given by
\begin{align}\label{FODTD}
G_E^{12}(x_1,x_2)\equiv\ave{\phi_1(x_1) \phi_2(x_2)}=h\int_{\pr
EAdS}K_{E\De_1}(\bx_1,\te_1;\bx)K_{E\De_2}(\bx_2,\te_2;\bx)d^d\bx,
\end{align}
where $K_{E\De}$ is the bulk to boundary propagator,
related to the bulk to bulk propagator by
\begin{equation}
K_{E\De}(\bx',\te';\bx)=lim_{\te\ra \pi/2}
\frac{2\De-d}{(\cos(\te))^\De}G_{E\De}(\te,\bx;\te',\bx').
\end{equation}

For simplicity we will restrict to the case where $x_2$ is at the
origin of the coordinate system, $\te_2=\tau_{E2}=0$, and $x_1$ also has
$\te_1=0$ (this is not the most general case, but for our purposes
it is enough to see that the commutator is non-vanishing for this
choice of pairs of points).
For points with $\te=0$, the bulk to boundary
propagator has the simple form:
\begin{align}\label{btbp}
K_{E\De}(\tau_E,\Om_{d-1},\te=0;\tau_E',\Om_{d-1}')=
C_K(\De)\cosh^{-\De}(\tau_E'-\tau_E),
\end{align}
with
\begin{align}
C_K(\De)=\frac{\Gamma(\De)(2\De-d)}{\pi^{\haf
d}R^{d-1}\Gamma(\De-\haf(d-2))2^{\De+1}}.
\end{align}

Plugging \eqref{btbp} in the expression for the propagator
correction \eqref{FODTD}, we get
\begin{align}\label{Edouble trace}
G_E^{12}(\te_1=0,\tau_{E1}=\tau_E;&\te_2=0,\tau_{E2}=0)=\cr
&hA\int_{-\infty}^{\infty}\cosh^{-\De_1}(\tau'-\tau_{E})
\cosh^{-\De_2}(\tau')d\tau',
\end{align}
where $A=V_{S^{d-1}}C_K(\De_1)C_K(\De_2)$ and $V_{S^{d-1}}$ is the
volume of $S^{d-1}$ coming from the integration over the $S^{d-1}$
on the boundary.

\begin{figure}[t]
\begin{center}
\includegraphics[height=8cm]{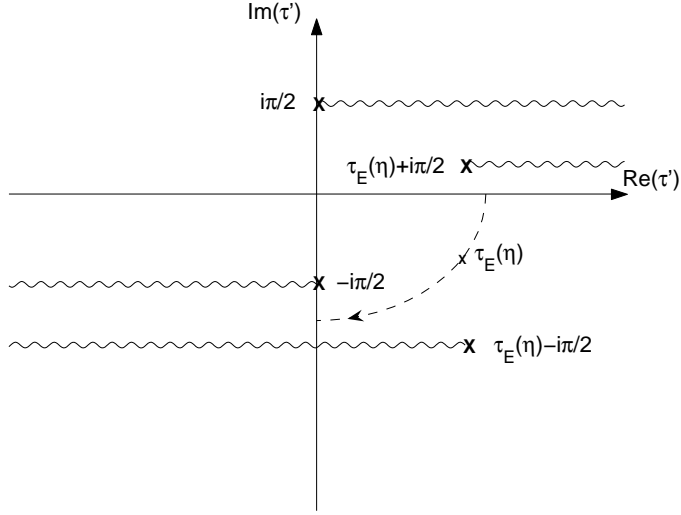}%
\caption{The branch cuts of the function
$\cosh^{-\De_1}(\tau'-\tau_E(\eta))\cosh^{-\De_2}(\tau')$ during
the $\tau_E$ rotation. The dashed curve shows the trajectory
followed by the point $\tau_E(\eta)$, and the branch cuts are drawn
for a particular point on this trajectory.\label{branch_mid}}
\end{center}
\end{figure}

To get the Lorentzian Feynman propagator we need to analytically rotate the
expression \eqref{Edouble trace} along the contour
\begin{align}
\tau_E(\eta)=e^{-i\eta}\tau,~~~~ \eta:0\ra \haf\pi.
\end{align}
The functions $\cosh^{-\De}(\tau)$ have singularities and (for
non-integer $\De$) branch cuts
in the complex plane; we discuss their form in detail in appendix
\ref{double_appendix}.
While rotating $\tau_E$, the contour of the $\tau'$ integration
needs to stay between the singularities at $\tau'=-i\haf\pi$
coming from $\cosh^{-\De_2}(\tau')$ and the singularity at
$\tau'=\tau_E(\eta)+i\haf\pi$ coming from
$\cosh^{-\De_1}(\tau'-\tau_E(\eta))$ (see figure
\ref{branch_mid}). Also note that the exponential falloff at large
$\tau'$ allows us to move the contour at infinity without
changing the value of the integral.

The way to perform this rotation depends on the value of $\tau$. We
analyze 2 cases :

Case 1) $\tau<\pi$ :
In this case we can choose the integration contour while rotating
$\tau_E$ to be the contour parallel to the real axis at
$i\frac{\trm{Im}(\tau_E(\eta))}{2}$, ending up at the contour
$\trm{Im}(\tau')=-\frac{i\tau}{2}$.
After completing the rotation we end up with the integral (see
figure \ref{contour1}):
\begin{figure}[t]
\begin{center}
\includegraphics[height=8cm]{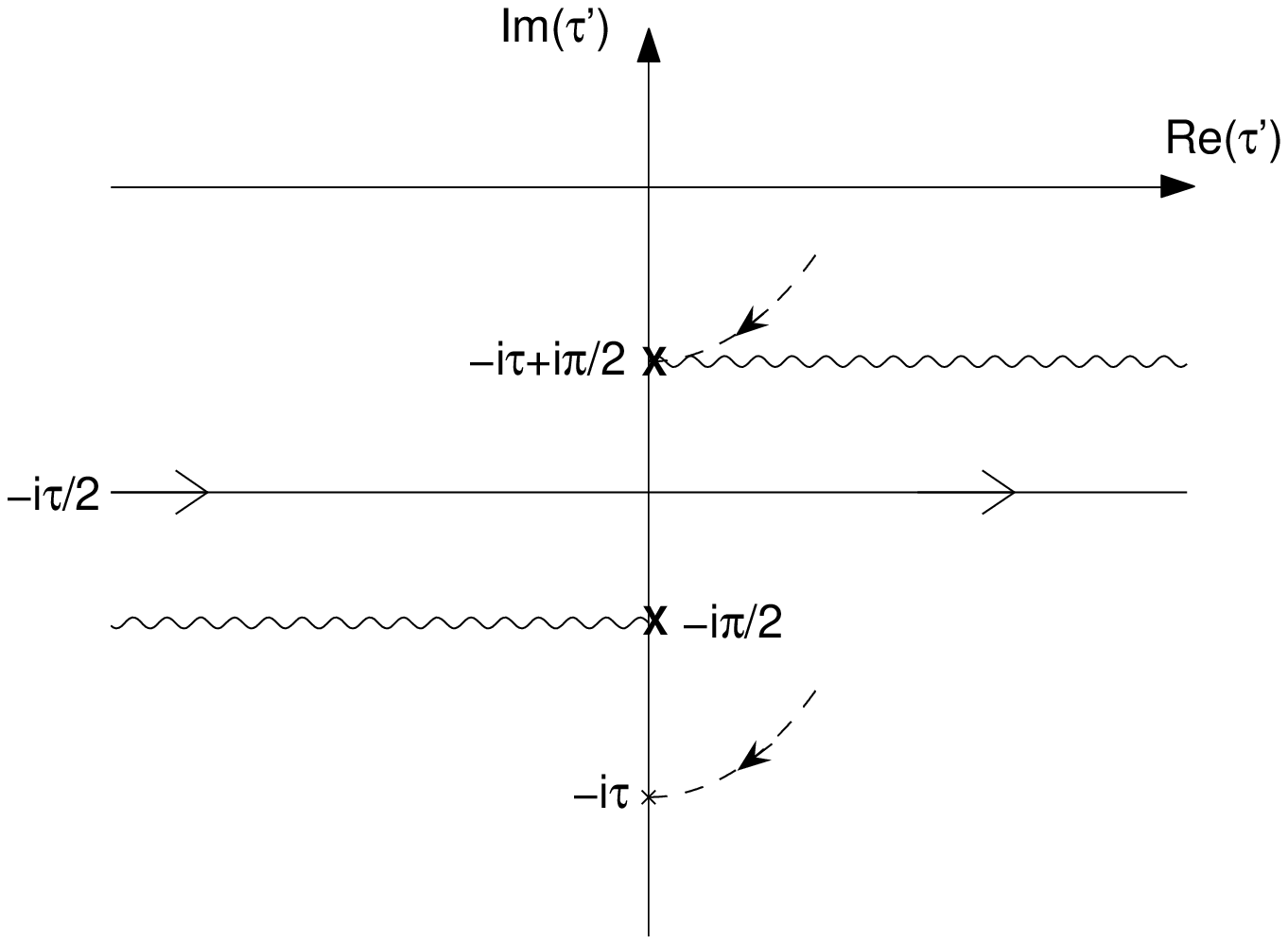}%
\caption{The $\tau'$ contour after $\tau_E$ rotation for
$0<\tau<\pi$. The relevant branch cuts of the function
$\cosh^{-\De_1}(\tau'+i\tau)\cosh^{-\De_2}(\tau')$ are shown on
the figure. The dashed curves show the last part of the trajectory
followed by the points $\tau_E(\eta)$ and $\tau_E(\eta)+i\pi/2$
during the rotation.\label{contour1}}
\end{center}
\end{figure}
\begin{align}
G^{12}(\te_1=0,&\tau_1=\tau;\te_2=0,\tau_2=0)=\cr&
hA\int_{-\infty}^{\infty}\cosh^{-\De_1}(\tau'+i\haf\tau)\cosh^{-\De_2}
(\tau'-i\haf\tau)d\tau',
\end{align}
the complex conjugate of which is (see appendix
\ref{double_appendix}):
\begin{align}
(G^{12})^*&=hA\int_{-\infty}^{\infty}\cosh^{-\De_1}(\tau'-i\haf\tau)
\cosh^{-\De_2}(\tau'+i\haf\tau)d\tau'=\cr
&=hA\int_{-\infty}^{\infty}\cosh^{-\De_1}(-\tau'+i\haf\tau)
\cosh^{-\De_2}(-\tau'-i\haf\tau)d\tau'=\cr
&=hA\int_{-\infty}^{\infty}\cosh^{-\De_1}(\tau'+i\haf\tau)
\cosh^{-\De_2}(\tau'-i\haf\tau)d\tau'=G^{12}.
\end{align}
So, the first order correction to the commutator of the two fields
vanishes in this case,
\begin{align}
i\trm{Im}(G^{12})=0.
\end{align}
This is consistent with our expectations, since in this case
$x_1$ is not in the `boundary affected' region.

Case 2) $\pi<\tau<2\pi$ :
\begin{figure}[t]
\begin{center}
\includegraphics[height=8cm]{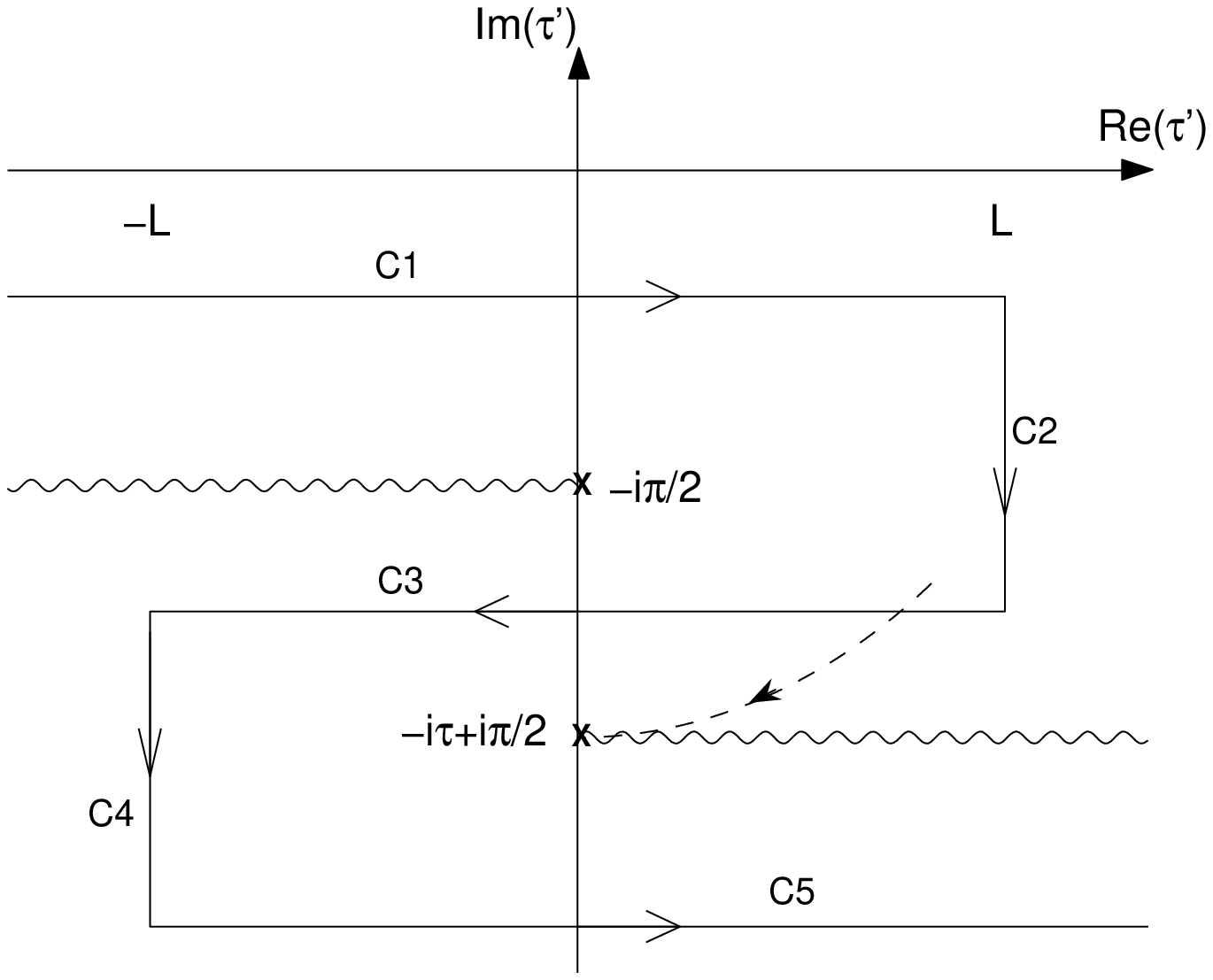}%
\caption{The $\tau'$ contour after $\tau_E$ rotation for
$\pi<\tau<2\pi$. The relevant branch cuts of the function
$\cosh^{-\De_1}(\tau'+i\tau)\cosh^{-\De_2}(\tau')$ are shown on
the figure. The dashed curve shows the last part of the trajectory
followed by the point $\tau_E(\eta)+i\pi/2$ during the
rotation.\label{contour2}}
\end{center}
\end{figure}
In this case, in order to avoid the poles at $\tau'=-i\haf\pi$ and
$\tau'=\tau_E(\eta)+i\haf\pi$, we need to deform the contour while
rotating $\tau$. After the rotation is complete we end up with the
contour shown in figure \ref{contour2}. Divide this contour into 5
straight contours: $C=C_1+C_2+C_3+C_4+C_5$ as shown in the figure,
\begin{align}
&C_1=(-\infty-i\haf\tau+i\haf\pi)\ra(L-i\haf\tau+i\haf\pi)\cr
&C_2=(L-i\haf\tau+i\haf\pi)\ra(L-i\haf\tau)\cr
&C_3=(L-i\haf\tau)\ra(-L-i\haf\tau)\cr &C_4=(-L-i\haf\tau)\ra
(-L-i\haf\tau-i\haf\pi)\cr
&C_5=(-L-i\haf\tau-i\haf\pi)\ra(\infty-i\haf\tau-i\haf\pi)
\end{align}
where $L$ is some real positive number. The specific location and
form of the contour were chosen for convenience.

By taking $L$ to infinity,
the contribution of the sections $C_2$ and $C_4$ goes to zero
and we end up with the union of
\begin{align}
&C_1=(-\infty-i\haf\tau+i\haf\pi)\ra(\infty-i\haf\tau+i\haf\pi)\cr
&C_3=(\infty-i\haf\tau)\ra(-\infty-i\haf\tau)\cr
&C_5=(-\infty-i\haf\tau-i\haf\pi)\ra(\infty-i\haf\tau-i\haf\pi)
\end{align}

It is not hard to see that (see
\eqref{C1},\eqref{C_3})
\begin{align}\label{relation}
&\int_{C_5}=e^{i\pi(\De_1+\De_2)}(\int_{C_1})^*,\cr
&\int_{C_3}=e^{2\pi i(\De_1+\De_2)}(\int_{C_3})^*,
\end{align}
where $\int_\gamma$ stands for
$\int_\gamma\cosh^{-\De_1}(\tau'+i\tau)\cosh^{-\De_2}(\tau')d\tau'$.
Thus, when $\De_1+\De_2=n$ for some integer even $n$, the sum of the
integrals is real and the commutator is zero. However, for odd $n$
or for non-integer $\De_1+\De_2$ it is not zero in general. In particular
there is a simple argument that it is not zero for $\tau \ra \pi^+$. From
\eqref{relation} we can deduce that the phase of $\int_{C_3}$ is
$\pi(\De_1+\De_2)$ (up to addition of $\pi k$),
$\int_{C_3}=\pm\abs{\int_{C_3}}e^{i\pi(\De_1+\De_2)}$. For general
$\De_1,\De_2$, for $\tau\ra\pi^+$, the integrals $\int_{C_1}$ and
$\int_{C_5}$ are finite, while the integral $\int_{C_3}$ goes to
infinity. Since for $\De_1+\De_2$ non-integer the phase of
$\int_{C_3}$ is not zero, the commutator which is its imaginary
part goes to infinity as well. In particular it is not zero.
This proves that the commutator of $\Phi_1$ and $\Phi_2$ does not
vanish for the pairs of points we are interested in after the
deformation (at least for specific values of $\tau$; we expect
similar arguments to apply for any $\tau > \pi$).

\subsection{Marginal deformation}
A special case where we can compute the effect of the deformation
\eqref{DTD} to all orders is the classically marginal case:
\begin{align}\label{d-De}
\De_1=d-\De_2\equiv\De.
\end{align}
Equation \eqref{d-De} means that
$m_1^2=m^2_2\equiv m^2$, the only difference between the fields in
the non-deformed case is their boundary condition.
The fact that $\phi_1$ and $\phi_2$ have the same mass allows us to
make the rotation:
\begin{align}\label{fields redifinition}
\rho_1=\inv{\sqrt{1+\hh^2}}(\phi_1+\hh\phi_2),
~~~\rho_2=\inv{\sqrt{1+\hh^2}}(-\hh\phi_1+\phi_2),
\end{align}
where $\hh=(2\De-d)h$. It is easy to show that if $\phi_1$ and $\phi_2$
satisfy the deformed boundary conditions \eqref{DTBC} then
the new fields, $\rho_1$ and $\rho_2$,
satisfy the same bulk equations as $\phi_1$ and $\phi_2$ but have
the undeformed boundary conditions, with $\rho_1$ behaving like a field dual
to an operator with dimension
$\De$ and $\rho_2$ with dimension $d-\De$. Since these are
decoupled fields their propagator is trivial and we can substitute
back to compute the propagator of $\phi_1$ and $\phi_2$ :
\begin{align}\label{fieldsresult}
G_E^{ij}=\inv{1+\hh^2}\left(\begin{array}{cc}G_{E\De}+\hh^2G_{E(d-\De)}&~~~~\hh
G_{E\De}-\hh G_{E(d-\De)}
\\\hh G_{E\De}-\hh G_{E(d-\De)}&~~~~
\hh^2G_{E\De}+G_{E(d-\De)} \end{array}\right).
\end{align}
To first order in $h$, there is only a correction to $G_E^{12}$,
\begin{align}
G_E^{12}=\hh(G_{E\De}-G_{E(d-\De)}),
\end{align}
which can be shown to agree with \eqref{FODTD}.

\section{Summary}\label{summary}

In this paper
we analyzed the effect of multi-trace deformations of the CFT on
the bulk to bulk propagators of the corresponding $AdS_{d+1}\times
M$ theory. We found that the resulting `bulk to bulk' commutators
of fields on $AdS_{d+1}\times M$ have a non-local property: they
do not vanish for pairs of points in $AdS_{d+1}\times M$ that are
connected by a space-like geodesic but not by a time-like geodesic
(we would expect a commutator of fields to vanish at such points
since the path integral over paths would be expected to be
dominated by the space-like geodesic; these points are connected
by a causal curve so the result does not violate causality). We
showed that in a standard theory (with only single-trace
deformations) the commutator of two scalar fields\footnote{We have
also proven this for vector fields, and we expect it to hold for
any pair of fields.} vanishes at any such pairs of points (at
least for odd-dimensional AdS spaces). However, this is no longer
true after deforming by multi-trace deformations (at least for the
specific deformation we considered, and for specific choices of
pairs of points, but we expect the result to be much more general).

\bigskip
\bigskip
\centerline{\bf Acknowledgements}

We would like to thank T. Kashti, A. Patir, D. Reichmann and E. Silverstein
for useful discussions.
The work of OA and MB was
supported in part by the Israel-U.S. Binational Science Foundation,
by the Israel Science Foundation (grant number 1399/04), by the
Braun-Roger-Siegl foundation, by the European network
HPRN-CT-2000-00122, by a grant from the G.I.F., the German-Israeli
Foundation for Scientific Research and Development, and by Minerva.

\appendix
\section{Propagators in AdS space}\label{propagator_in_AdS}

In this appendix we will review some properties of the Feynman
propagators in AdS space. The main result used in the text is the
relation between the Feynman propagators in different space-time
dimensions \eqref{Aprop rel}.

\subsection{Propagator in Euclidean signature}

Instead of working with the metric \eqref{metrictauEtheta}, we define
a new radial coordinate $l$ by the geodesic distance from the origin,
and then we can write the $EAdS_{d+1}$ metric as
\begin{align}
ds^2=R^2(dl^2+\sinh^2(l)d\Om_d^2).
\end{align}
The Laplacian in $EAdS_{d+1}$ in these coordinates is
\begin{align}
\nabla^2\phi&=\inv{R^2}(R\sinh l)^{-d}\pr_l((R\sinh
l)^d\pr_l\phi)+\inv{R^2\sinh^2 l}\nabla^2_d\phi=\cr
&=\inv{R^2}\left(\pr_{l}^2+d\coth(l)\pr_l+\inv{\sinh^2(l)}\nabla^2_d\right)
\phi,
\end{align}
where $\nabla^2_d$ is the Laplacian on $S^d$.
The Klein-Gordon (KG) equation, $(\nabla^2 -m^2)\phi=0$, in the
case of spherical symmetry is:
\begin{align}
\left(\pr_{l}^2+d\coth(l)\pr_l\right)\phi=R^2m^2\phi,
\end{align}
or:
\begin{align}\label{propsimp}
f''+d\frac cs f'=R^2m^2f,
\end{align}
where
\begin{align}
&\phi=f(l),\cr &s\equiv \sinh(l),c\equiv \cosh(l),\cr &f'\equiv \pr_lf \soo
c'=s,~~~s'=c.
\end{align}
Suppose we have a solution $f$ to equation (\ref{propsimp}).
Define
\begin{align}
g=f'/s \oso f'=sg\soo f''=cg+sg'.
\end{align}
Plug this in (\ref{propsimp}):
\begin{align}
&sg'+(d+1)cg=R^2m^2f\soo\cr &sg''+(d+1)sg+(d+2)cg'=R^2m^2sg\soo\cr
&g''+(d+2)\frac cs g'=R^2(m^2-\frac{(d+1)}{R^2})g.
\end{align}
So, $g$ is a solution to (\ref{propsimp}) in $EAdS_{d+3}$ with mass
\begin{align}\label{mass relation}
m_{d+2}^2=m_d^2-\frac{(d+1)}{R^2}.
\end{align}
Note that (\ref{mass relation}) means (using \eqref{de to m}):
\begin{align}
\De_{d+2}-\haf(d+2)=\sqrt{(\haf(d+2))^2+m_{d+2}^2R^2}=\sqrt{(\haf
d)^2+R^2m_d^2}=\De_d-\haf d.
\end{align}

So, if $\phi_d$ is a spherically symmetric solution to the KG
equation in $EAdS_{d+1}$ corresponding to dimension $\De_d$, then
$\phi_{d+2}$ defined as:
\begin{align}\label{sol rel}
\phi_{d+2}(l)=\inv{\sinh(l)}\frac{d}{dl}\phi_d
\end{align}
is a spherically symmetric solution to the KG equation in
$EAdS_{d+3}$ corresponding to a conformal dimension
$\De_{d+2}=\De_d+1$. The propagator is the solution to the KG
equation for $l>0$ with the large $l$ behavior:
\begin{align}
\phi_d(l)\propto e^{-\De_d l}.
\end{align}
By using \eqref{sol rel} we see that $\phi_{d+2}$ has the same
behavior as $\phi_d$, with $\De_{d+2}$. We conclude that the
propagators in dimensions $d+1$ and $d+3$ are related:
\begin{align}\label{prop rel}
&G_{EAdS_{d+3}}(l,\De)\propto\inv{\sinh(l)}\frac{d}{dl}G_{EAdS_{d+1}}(l,\De-1).
\end{align}

To check that this works we can calculate the propagator for some
low dimensions. For $d=0$, \eqref{propsimp} has the
solutions $e^{\pm mRl}=e^{\De_\pm l}=e^{\pm \De l}$.
For any even $d$, we can get
the relevant solution by recursion using \eqref{prop rel}.
For the lowest odd dimension
$AdS_3$ we get
\begin{align}
G_{EAdS_3}(l)\propto\frac{e^{-(\De-1)l}}{\sinh(l)}.
\end{align}
We can calculate the normalization factor by the requirement that
the propagator should approach the flat space propagator for
$Rl<<R,1/m$. We get:
\begin{align}\label{prop35}
G_{EAdS_3}(l)=-\inv{4\pi R}\frac{e^{-(\De-1)l}}{\sinh(l)},
\end{align}
in agreement with the literature \cite{Inami:1985wu}.

\subsection{Propagator in Lorentzian signature}

It will be convenient to
work in global coordinates $(\rho,\tau,\Om_{d-1})$ with the metric
\begin{equation} \label{metrictaurho}
ds^2=R^2(-\cosh^2 \rho~d\tau^2+d\rho^2+\sinh^2\rho~d\Omega_{d-1}^2).
\end{equation}
The propagator in Lorentzian signature is expressed as an analytic
continuation of the propagator in $EAdS$, as a function of
$\tau_E$, back to real time $\tau=-i\tau_E$. This means rotating
$\tau_E$ along the contour:
\begin{align}
\tau_E(\eta)=e^{-i\eta}\tau,~~~~ \eta:0\ra \haf\pi.
\end{align}

For $\rho=0$, $l=\tau_E$, the expressions \eqref{prop35} are analytic
functions of $\tau_E$ without any branch cuts, so the rotation can
be done trivially and we get the Feynman propagator
\begin{align}
G_{AdS_3}(x,0)=i\inv{4\pi R}\frac{e^{i(\De-1)\tau}}{\sin\tau}.
\end{align}
 Since the propagator is invariant under the isometries of AdS, it must
be a function of the geodesic proper time $\mu$ between the two points
(this is true for pairs of points that have a timelike geodesic
connecting them, which are the pairs that interest us; assume that the
first point has a larger $\tau$ so that the time ordering has no
effect). Thus we get:
\begin{align}\label{prop3}
&G_{AdS_3}(\mu;m^2)=i\inv{4\pi R}\frac{e^{i(\De-1)\frac\mu
R}}{\sin\frac\mu R}=i\inv{4\pi
R}\frac{e^{i\mu\sqrt{m^2+\inv{R^2}}}}{\sin\frac\mu R}.
\end{align}
Similarly, by analytically continuing \eqref{prop rel} we get
\begin{align}\label{Aprop rel}
&G_{AdS_{d+3}}(\mu,m^2)\propto\inv{\sin(\frac\mu
R)}\frac{d}{d\mu} G_{AdS_{d+1}}(\mu,m^2+\frac{d+1}{R^2}).
\end{align}

\section{Properties of the function
$\cosh^{-\De_1}(\tau'-\tau)\cosh^{-\De_2}(\tau')$}\label{double_appendix}

\subsection{Properties of $\cosh^\al(z)$}

\begin{figure}[t]
\begin{center}
\includegraphics[height=8cm]{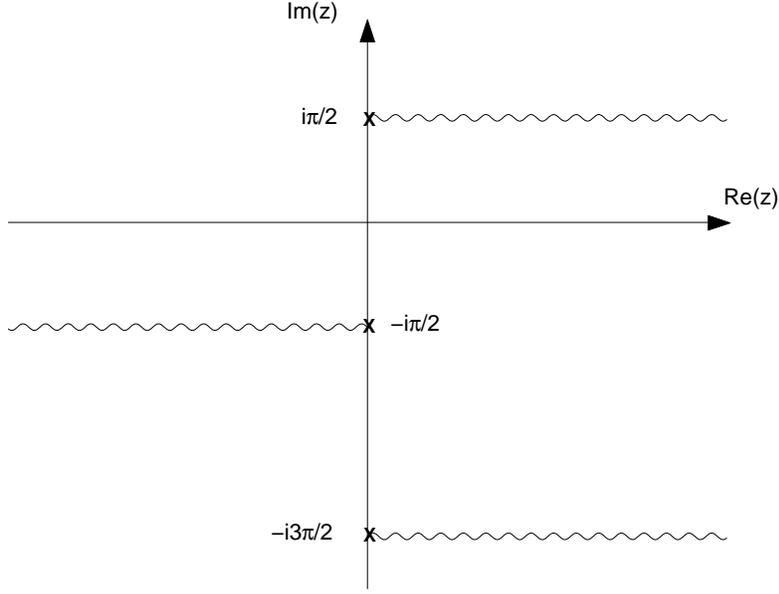}%
\caption{The branch cuts of the function
$\cosh^\al(z)$\label{branch1}}
\end{center}
\end{figure}

The function $\cosh(z)$ has zeros at $z=i\haf\pi+i\pi k$ for integer $k$.
Define the
branch cut of the function $z^{\al}$ to be at $(0+0i,0+i\infty)$.
This means we write
\begin{align} \label{zdedef}
z^{\al}\equiv
\abs{z}^{\al}e^{i\sig\al},~~\textrm{where}~~z=\abs{z}e^{i\sig}~~
\textrm{with}~~-\frac32
\pi<\sig\leq\haf\pi.
\end{align}
With this choice the function $\cosh^\al(z)$ has branch cuts along the
half-lines
$(\trm{Im}(z)=\haf\pi+2\pi k, \trm{Re}(z)>0)$  and
$(\trm{Im}(z)=-\haf\pi+2\pi k, \trm{Re}(z)<0)$ (see figure
\ref{branch1} for the first few branch cuts).

Note that with the definition \eqref{zdedef}
\begin{align}
&(z^{\al})^*=\left\{\begin{array}{cc} (z^*)^\al &
-\haf\pi\leq\sig\leq\haf\pi\\
e^{2\pi i\al}(z^*)^\al & -\frac32\pi<\sig<-\haf\pi\\
\end{array} \right.\cr
&(-z)^\al=\left\{\begin{array}{cc} e^{-i\pi\al}z^\al &
-\haf\pi\leq\sig\leq\haf\pi\\
e^{i\pi\al}z^\al & -\frac32\pi<\sig<-\haf\pi\\
\end{array} \right.
\end{align}
The function $\cosh(z)$ with $z=\trm{Re}(z)+i\trm{Im}(z)$ obeys
\begin{align}
\cosh(z)=\cosh(\trm{Re}(z))\cos(\trm{Im}(z))+
i\sinh(\trm{Re}(z))\sin(\trm{Im}(z)).
\end{align}
Note that for
\begin{align}
\cosh(z)=\abs{\cosh(z)}e^{i\sig}~~\textrm{with}~~-\frac32
\pi<\sig\leq\haf\pi
\end{align}
we find that (up to $z\ra z+2\pi i k$)
\begin{align}
&-\haf\pi\leq\sig\leq\haf\pi \oso
-\haf\pi\leq\trm{Im}(z)\leq\haf\pi \cr &-\frac32\pi<\sig<-\haf\pi
\oso -\frac32\pi<\trm{Im}(z)<-\haf\pi,
\end{align}
so
\begin{align}
&(\cosh(z)^{\al})^*=\left\{\begin{array}{cc} \cosh^\al(z^*) &
-\haf\pi\leq\trm{Im}(z)\leq\haf\pi \\
e^{2\pi i\al}\cosh^\al(z^*)& -\frac32\pi<\trm{Im}(z)<-\haf\pi\\
\end{array} \right.\cr
&(-\cosh(z))^\al=\left\{\begin{array}{cc} e^{-i\pi\al}\cosh^\al(z)
&
-\haf\pi\leq\trm{Im}(z)\leq\haf\pi\\
e^{i\pi\al}\cosh^\al(z) & -\frac32\pi<\trm{Im}(z)<-\haf\pi.\\
\end{array} \right.
\end{align}

To summarize, the poles and branch cuts of $\cosh^\al(z)$ are :
\begin{align}
\cosh(z)&=0\oso \trm{Re}(z)=0,\trm{Im}(z)=\haf\pi+\pi k,\cr
\cosh(z)&\in (0+i0,0+i\infty)\oso\cr & (\trm{Im}(z)=\haf\pi+2\pi
k,~\trm{Re}(z)>0)~ \textrm{or} ~(\trm{Im}(z)=-\haf\pi+2\pi
k,~\trm{Re}(z)<0).
\end{align}

\subsection{Derivation of equation (4.14)}

Divide each contour into two parts (-infinity to zero and zero to
infinity) :
\begin{align}
&C_{1a}=(-\infty-i\haf\tau+i\haf\pi)\ra(0-i\haf\tau+i\haf\pi)\cr
&C_{1b}=(0-i\haf\tau+i\haf\pi)\ra(\infty-i\haf\tau+i\haf\pi)\cr
&C_{3a}=(\infty-i\haf\tau)\ra(0-i\haf\tau)~~~
C_{3b}=(0-i\haf\tau)\ra(-\infty-i\haf\tau)\cr
&C_{5a}=(-\infty-i\haf\tau-i\haf\pi)\ra(0-i\haf\tau-i\haf\pi)\cr
&C_{5b}=(0-i\haf\tau-i\haf\pi)\ra(\infty-i\haf\tau-i\haf\pi)
\end{align}
Note that:
\begin{align}
&(\int_{C_{1a}})^*=e^{-2\pi
i\De_1}\int_{-\infty}^0\cosh^{-\De_1}(\tau'-i\haf\tau-i\haf\pi)
\cosh^{-\De_2}(\tau'+i\haf\tau-i\haf\pi)d\tau'=\cr
&=e^{-2\pi
i\De_1}\int^{\infty}_0\cosh^{-\De_1}(\tau'+i\haf\tau+i\haf\pi)
\cosh^{-\De_2}(\tau'-i\haf\tau+i\haf\pi)d\tau'=\cr
&=e^{-2\pi i\De_1}\int_{C_{1b}}=\cr &=e^{-2\pi
i\De_1}e^{+i\pi\De_1}e^{-i\pi\De_2}\int^{\infty}_0\cosh^{-\De_1}
(\tau'+i\haf\tau-i\haf\pi)\cosh^{-\De_2}(\tau'-i\haf\tau-i\haf\pi)d\tau'=\cr
&=e^{-i\pi (\De_1+\De_2)}\int_{C_{5b}},
\end{align}
where in the last line, we used the relation $\cosh(z\pm i\pi)=-\cosh(z)$.

Using similar relations we find:
\begin{align}
&\int_{C_{5a}}=e^{i\pi(\De_2-\De_1)}\int_{C_{1a}}=
e^{i\pi(\De_1+\De_2)}(\int_{C_{1b}})^*\cr
&\int_{C_{3b}}=e^{2\pi i(\De_1+\De_2)}(\int_{C_{3a}})^*.
\end{align}
Thus,
\begin{align}\label{C1}
\int_{C_5}&=\int_{C_{5a}}+\int_{C_{5b}}=
e^{i\pi(\De_1+\De_2)}(\int_{C_{1b}})^*+
e^{i\pi(\De_1+\De_2)}(\int_{C_{1a}})^*=\cr
&=e^{i\pi(\De_1+\De_2)}(\int_{C_1})^*
\end{align}
and
\begin{align}\label{C_3}
\int_{C_3}&=\int_{C_{3a}}+\int_{C_{3b}}=e^{2\pi
i(\De_1+\De_2)}(\int_{C_{3b}})^*+e^{2\pi
i(\De_1+\De_2)}(\int_{C_{3a}})^*=\cr &=e^{2\pi
i(\De_1+\De_2)}(\int_{C_3})^*.
\end{align}

\end{document}